\documentclass[conference]{IEEEtran}
\usepackage{cite}
\usepackage{amsmath,amssymb,amsfonts}
\usepackage{graphicx}
\usepackage{textcomp}
\usepackage{xcolor}
\usepackage{fancyhdr}
\usepackage[hyphens]{url}
\usepackage{fancyhdr}
\usepackage{comment}
\usepackage{algorithm}
\usepackage[noend]{algpseudocode} 
\algrenewcommand\textproc{}
\usepackage{placeins}
\usepackage{verbatim}
\usepackage{tikz}

\usepackage{graphicx}
\usepackage{subfig}

\def\BibTeX{{\rm B\kern-.05em{\sc i\kern-.025em b}\kern-.08em
    T\kern-.1667em\lower.7ex\hbox{E}\kern-.125emX}}

\title{PartitionedVC: Partitioned External Memory Graph Analytics Framework for SSDs}

\begin{document}
\author{\IEEEauthorblockN{Kiran Kumar Matam}
kmatam@usc.edu\\
\IEEEauthorblockA{Computer Science Department \\
University Of Southern California\\
}
\and
\IEEEauthorblockN{Hanieh Hashemi}
hashemis@usc.edu\\
\IEEEauthorblockA{Electrical and Computer Engineering \\
University Of Southern California\\
}
\and
\IEEEauthorblockN{Murali Annavaram}
annavara@usc.edu\\
\IEEEauthorblockA{Electrical and Computer Engineering \\
University Of Southern California\\
}
}
\maketitle
\pagestyle{plain}

\begin{abstract}
Graph analytics are at the heart of a broad range of applications such as drug discovery, page ranking, transportation systems, and recommendation systems. When graph size exceeds memory size, out-of-core graph processing is needed. For the widely used external memory graph processing systems, accessing storage becomes the bottleneck. We make the observation that nearly all graph algorithms have a dynamically varying number of active vertices that must be processed in each iteration. However, existing graph processing frameworks, such as GraphChi, load the entire graph in each iteration even if a small fraction of the graph is active. This limitation is due to the structure of the data storage used by these systems. In this work, we propose to use a compressed sparse row (CSR) based graph storage that is more amenable for selectively loading only a few active vertices in each iteration. But CSR based graph processing suffers from random update propagation to many target vertices. To solve this challenge, we propose to use a multi-log update mechanism that logs updates separately, rather than directly update the active edges in a graph. Our proposed multi-log system maintains a separate log per each vertex interval. This separation enables us to efficiently process each vertex interval by just loading the corresponding log. Further, while accessing SSD pages with fewer active vertex data, we reduce the read amplification due to the page granular accesses in SSD by logging the active vertex data in the current iteration and efficiently reading the log in the next iteration. Over the current state of the art out-of-core graph processing framework, our evaluation results show that the PartitionedVC framework improves performance by up to $17.84\times$, $1.19\times$, $1.65\times$, $1.38\times$, $3.15\times$, and $6.00\times$ for the widely used breadth-first search, pagerank, community detection, graph coloring, maximal independent set, and random-walk applications, respectively.

\end{abstract}

%
%


\vspace{-2mm}
\section{Introduction}
Graph analytics are at the heart of a broad range of applications. The size of the graphs in many of these domains exceeds the capacity of main memory. Hence, many out-of-core (also called external memory) graph processing systems have been proposed. These systems primarily operate on graphs by splitting the graph into chunks and operating on each chunk that fits in main memory. For instance, GraphChi~\cite{graphchi} and many follow on research papers~\cite{ai2018clip, zhang2018wonderland,vora2016load}, format graphs into shards, where each shard has a unique data organization that minimizes the number of random disk accesses needed to process one chunk of the graph at a time. 

In terms of programming, many popular graph processing systems use vertex-centric programming paradigm. This computational paradigm uses bulk synchronous parallel (BSP) processing where each vertex is processed at most once during a single superstep, which may generate new updates to their connected vertices. The graph is then iteratively processed in the following superstep. A vertex can modify its state or generate updates to another vertex, or even mutate the topology of the graph in each superstep. 

The repeated supersteps of execution in vertex-centric programming, however, cause a major data access hurdle for shard-based graph frameworks. We show later in this work that as supersteps progress, the number of active vertices that receive messages, and hence must process those messages, continuously shrinks. Shard-based graph frameworks are unable to limit their accesses to a limited number of active vertices. In each superstep, they still have to read whole shards into memory to access the active vertices that are embedded in these shards. 

It is this observation that active vertex set shrinks with each superstep that our work exploits as a first step. Given that shard-based graph formats do not support accessing just the active vertex set, we rely on an efficient compressed sparse row (CSR) representation of the graph to achieve this goal. 
Since CSR format can either place incoming or outgoing edges in contiguous locations but not both, the problem of random access traffic is one of the drawbacks of CSR format (more details in the next section).  
To resolve this challenge, we design a novel multi-log graph processing paradigm that logs all the outgoing messages into a collection of logs, where each log is associated with an interval of vertices. Logging the messages, as opposed to embedding the messages as metadata into each edge, decouples the message placement from the organization of edges in the CSR  format, thereby resolving the random access hurdle. 

The last challenge tackled by this work is that when using log-based message processing, each superstep still has to read the edge lists from storage. As we show later in our analysis, many of the edge list pages accessed in SSD contain edges that are from inactive vertices. Hence, to improve edge list access efficiency, we optimize the multi-log design by logging the edge lists of the vertex that is likely to be active in the next superstep. We use a simple history-based prediction to determine whether a vertex will be active in the next superstep.

Compared to prior approaches that also rely on log-based graph processing~\cite{GraFBoost,elyasi2019large}, our approach differs in two important ways. Most log-based graph processing frameworks use a single log and \textit{merge} the messages bound to a single destination vertex to optimize log access latency. However, many important graph algorithms, such as community detection \cite{FLP_implementation}, graph coloring \cite{gonzalez2012powergraph}, maximal independent set \cite{malewicz2010pregel} must process all the messages \textit{individually}. Hence, approaches that merge messages do not permit the execution of many important classes of graph algorithms. However, preserving all the messages without any merge process leads to extremely large log size thereby creating significant log processing overhead at the beginning of each superstep \cite{GraFBoost}. Inspired by these challenges, we design a novel multi-log paradigm that allows efficient access to smaller logs, based on the vertex interval that is currently being processed. By preserving the messages without merging, we enable all graph algorithms to be executed within our framework, thereby supporting the generality of GraphChi-based frameworks~\cite{graphchi,ai2018clip,zhang2018wonderland,vora2016load}. At the same time, we use CSR formatted graphs augmented with multi-log structures to access the active vertex set, thereby drastically reducing the number of unwanted page accesses to the storage.

Our main contributions in this work are:

\begin{itemize}
    \item We propose an efficient external memory graph analytics system, called PartitionedVC (multi-log with vertex-centric generality), which uses a combination of CSR graph format, and message logging to enable efficient processing of large graphs that do not fit in main memory.  Unlike shard-based graph structures, CSR format enables accessing only the pages containing active vertices in a graph. This capability is a key requirement for efficient graph processing since the set of active vertices shrink significantly with each successive superstep. Hence, CSR format reduces read amplification. 
    
    \item  To efficiently access all the updates in a superstep, PartitionedVC partitions the graph into multiple vertex intervals for processing. All the outgoing messages are placed into multiple logs indexed by the destination vertex interval. All the updates generated by one vertex interval to other vertex interval are stored in their corresponding log. When an interval of vertices is scheduled for processing, all the updates it needs are located in a single log. We use SSD's capability for providing parallel writes to multiple channels for concurrently handling multiple vertex interval logs with only small buffers on the host side. 
    
    \item Even with message logging, PartitionedVC must still access the outgoing edge lists of each active vertex to send messages for the next superstep.  Our analysis of edge list accesses showed that many pages accesses contain edge lists of inactive vertices interspersed with active vertices. To further reduce the read amplification associated with edge lists, PartitionedVC logs the edge list of any potential active vertex in current iteration so that this edge list can be read more efficiently from the log, rather than reading a whole page of unneeded edge lists.

\end{itemize}


\vspace{-2mm}
\section{Background and Motivation}
\label{PartitionedVC:sec:BackgroundAndMotivation}

\vspace{-1mm}
\subsection{Out-of-core graph processing}
\label{PartitionedVC:sec:graph_storage}

In the out-of-core graph processing context, graphs sizes are considered to be large when compared to the main memory size but can fit in the storage size of current SSDs (in Terabytes). GraphChi \cite{graphchi} is a representative out-of-core vertex-centric programming system and many further works built on top of this basic framework~\cite{ai2018clip, zhang2018wonderland,vora2016load}. We will thus describe the GraphChi graph formats and describe the challenges with shard-based graph processing.

\begin{figure}
	\centering
	\subfloat[CSR format representation for the example graph]{
		\includegraphics[width=0.96\linewidth]{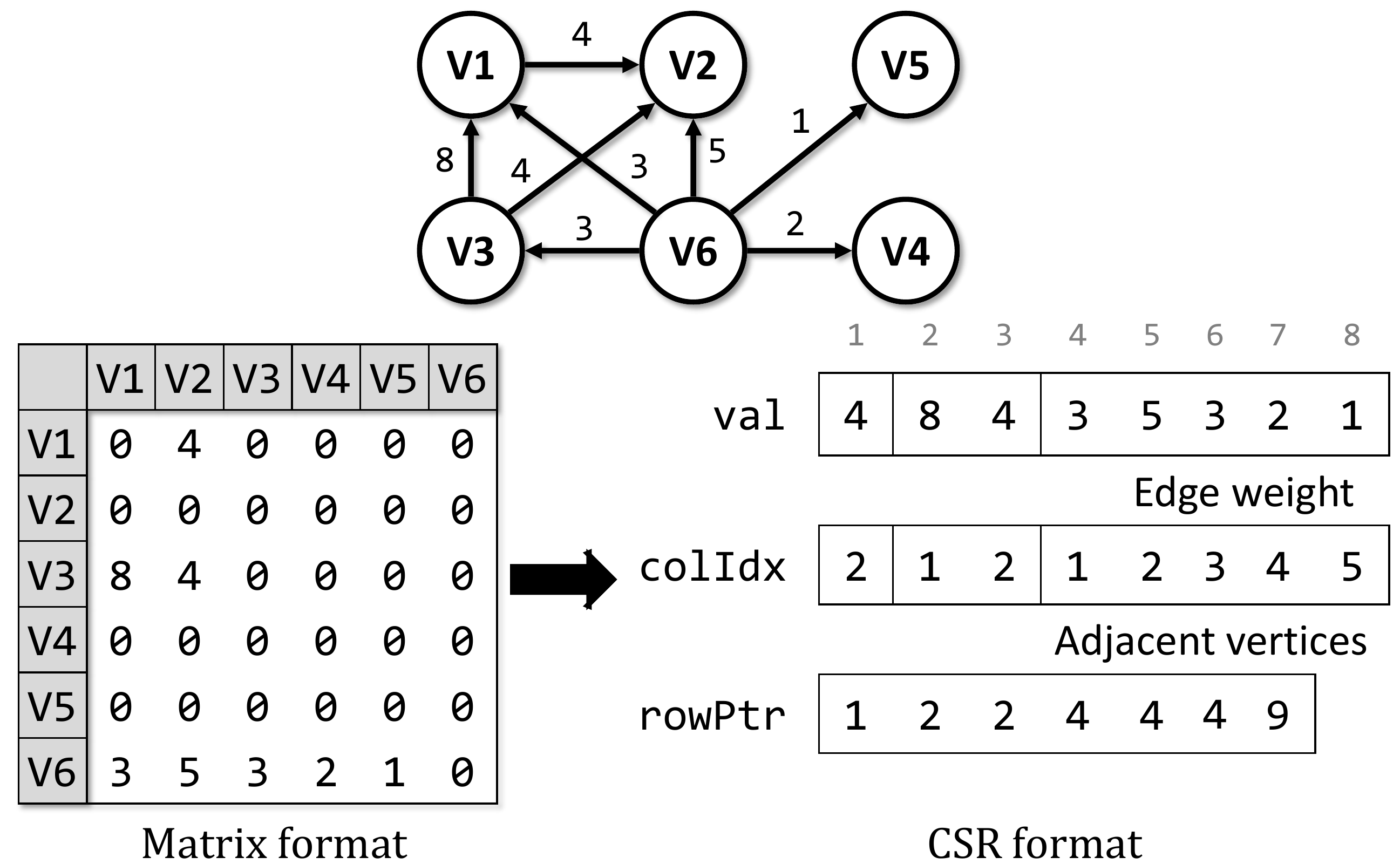}
		\label{fig:csr_format}
	}\\
	\subfloat[GraphChi shard structure for the example graph]{
		\includegraphics[width=0.96\linewidth]{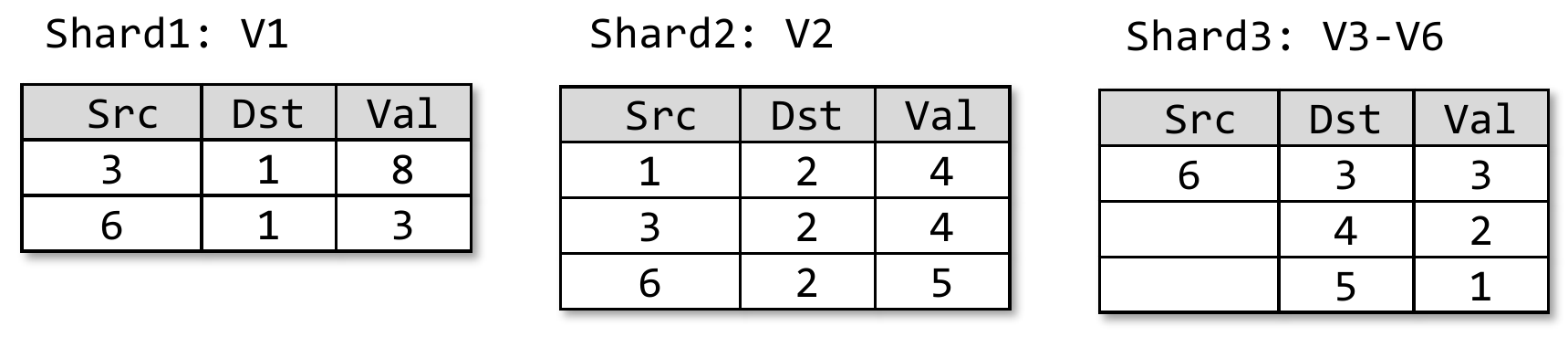}
		\label{fig:graphchi_format}
	}
	\vskip -2mm
	\caption{Graph storage formats}
	\vskip -4mm
	\label{fig:graph_formats}
\end{figure}

GraphChi partitions the graph into several vertex intervals, and stores all the incoming edges to a vertex interval as a shard. Figure \ref{fig:graphchi_format} shows the shard structure for an illustrative graph shown in Figure \ref{fig:csr_format}. For instance, shard1 stores all the incoming edges of vertex interval $V1$, shard2 stores $V2$'s incoming edges, and shard3 stores incoming edges of all the vertices in the interval $V3-V6$. While incoming edges are closely packed in a shard, the outgoing edges of a vertex are dispersed across other shards. In this example, the outgoing edges of $V6$ are dispersed across shard1, shard2, and shard3.   Another unique property of shard organization is that each shard stores all its in-edges sorted by source vertex.    

GraphChi relies on this shard organization to process vertices in intervals. It first loads into memory a shard corresponding to one vertex interval, as well as all the outgoing edges of those vertices that may be stored across multiple shards. Updates generated during processing are directly passed to the target vertices through the out-going edges in other shards in the memory. Once the processing for a vertex interval in a superstep is finished, its corresponding shard and its out-going edges in other shards are written back to the disk. 

GraphChi relies on sequential accesses to disk data and minimizes random accesses. However, in the following superstep, only a subset of vertices may become active. 
Due to shard organization above, even if a single vertex is active within a vertex interval the entire shard must be loaded since the in-edges for that vertex may be dispersed throughout the shard. 
For instance, if any of the $V3, V4, V5$ or $V6$ is active, the entire shard3 must be loaded. Loading a shard may be avoided only if all the vertices in the associated vertex interval are not active. However, in real-world graphs, the vertex intervals typically span tens of thousands of vertices, and during each superstep the probability of a single vertex being active in a given interval is very high. As a result, GraphChi in practice ends up loading all the shards in every superstep independent of the number of active vertices in that superstep.
\vspace{-2mm}
\subsection{Shrinking size of active vertices}
To quantify the number of superfluous page loads that must be performed by shard-based graph processing frameworks we measured the active vertex and the active edge counts in each superstep while running graph coloring application described in section \ref{PartitionedVC:sec:applications} over the datasets shown in Table \ref{PartitionedVC:table:graph_dataset}.  For this application, we ran a maximum of 15 supersteps. 

\begin{figure}
	\centering
	\includegraphics[width=\linewidth]{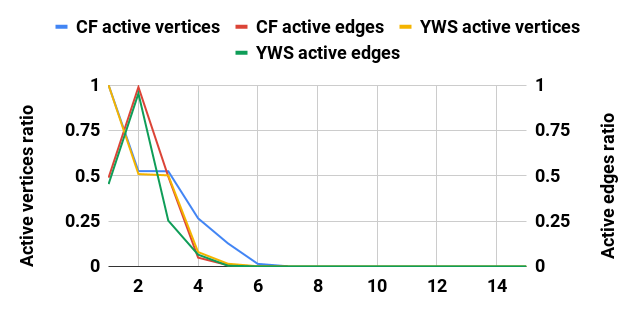}
	\vskip -3mm
	\caption{Active vertices and edges over supersteps}
	\vskip -4mm
	\label{PartitionedVC:fig:motivation}
\end{figure}

Figure \ref{PartitionedVC:fig:motivation} shows the active vertices and active edges count as a fraction of the total vertices and edges in the graph, respectively. The x-axis indicates the superstep number, the major y-axis shows the ratio of active vertices divided by total vertices, and the minor y-axis shows the number of active edges (updates sent over an edge) divided by the total number of edges in the graph. The fraction of active vertices and active edges shrink dramatically as supersteps progress. However, at the granularity of a shard, even the few active vertices lead to loading many shards since the active vertices are spread across the shards.

\vspace{-2mm}
\section{CSR format in the era of SSDs}
Given the overheads of loading is a significant fraction of the shards even with few active vertices in GraphChi, we evaluated a compressed sparse row (CSR) format for graph processing. CSR format has the desirable property that one may load just the active vertex information more efficiently. Large graphs tend to be sparsely connected. 

CSR format takes the adjacency matrix representation of a graph and compresses it using three vectors. The value vector, \textit{val}, stores all the non-zero values from each column sequentially. The column index vector, \textit{colIdx}, stores the column index of each element in the \textit{val} vector. The row pointer vector, \textit{rowPtr}, stores the starting index of each row (adjacency matrix row) in the \textit{val} vector. CSR format representation for the example graph is shown in Figure~\ref{fig:csr_format}. The edge weights on the graph are stored in \textit{val} vector, and the adjacent outgoing vertices are stored in \textit{colIdx} vector. To access adjacent outgoing vertices associated with a vertex in CSR graph storage format, we first need to access the \textit{rowPtr} vector to get the starting index in the \textit{colIdx} vector where the adjacent vertices associated with the vertex are stored in a contiguous fashion.

In CSR format, as all the outgoing edges connected to a vertex are stored in a contiguous location, while accessing the adjacency information for the active vertices CSR format is suitable for minimizing the number of pages accessed in an SSD and reducing the read amplification.

\begin{figure}
	\centering
	\includegraphics[width=\linewidth,height=4cm]{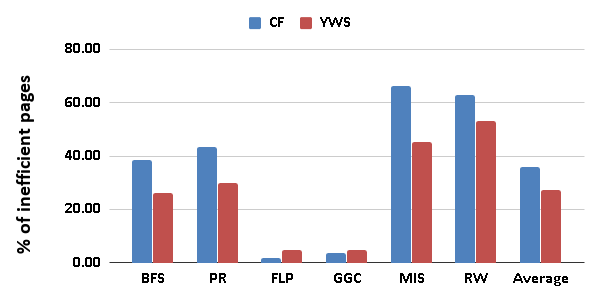}
	\vskip -3mm
	\caption{Accessed graph pages with less than 10\% of utilization}
	\vskip -4mm
	\label{PartitionedVC:fig:edge_list_motivation}
\end{figure}
\vspace{-2mm}
\subsection{Challenges for graph processing with a CSR format}
While CSR format looks appealing for accessing active vertices, it suffers one fundamental challenge. One can either maintain an in-edge list in the colIdx vector or the out-edge list, but not both (due to coherency issues with having the same information in two different vectors). Consider the case that adjacency list stores only in-edges and during the superstep all the updates on the out-edges generate many random accesses to the adjacency lists to extract the out-edge information. Similarly, if the adjacency list stores only out-edges then at the beginning of a superstep each vertex must parse many adjacency lists to extract all the incoming messages. 


\vspace{-2mm}
\section{PartitionedVC: Three Key Insights}
In this section we layout three key insights that lead to the design of the PartitionedVC. 
\vspace{-2mm}
\subsection{Avoid random write overhead with logging}
To avoid the random access problem with the CSR format, we exploit the first observation. Namely, updates to the out-edges do not need to be propagated by accessing the edge list.  Instead, these updates can be simply logged separately. The logged message has to store the destination vertex alongside the edge update so that the log can track which destination vertex this message is bound for. Hence, logging requires the addition of the destination vertex field to the message.  Logging the messages has two benefits. First, like any log structure, all the message writes occur sequentially, enabling an SSD to efficiently write the messages to storage for processing later.  Second, one can decouple the message logs from the CSR storage format inefficiencies; namely random accesses to either the in-edge or out-edge lists as we described earlier. Thus, we propose to log the updates sent between the vertices, instead of directly updating the edge values in the CSR format. Note that logging has been proposed in prior graph processing works \cite{GraFBoost,elyasi2019large}. But as we explain below, prior logging schemes did not exploit the second key observation we make in this work.

%
Prior works maintain a single log for all the updates in a superstep \cite{GraFBoost}. But the start of the next superstep the entire log must be parsed to find all the messages bound to a given destination vertex. Almost all prior works employ sorting of the log (based on a vertex ID) to efficiently extract all the messages bound for that vertex. In the worst case, the number of updates sent between the vertices may be proportional to the number of edges. Hence, the log itself must be stored in SSD. Even if a small fraction of edges receives an update, they may still overwhelm the host memory, given the size of the graphs.  Thus, one has to do external sorting of the message log. Prior works either built a custom accelerator \cite{GraFBoost} or proposed  approaches to reduce the sorting overhead \cite{elyasi2019large}. 
\vspace{-2mm}
\subsection{Eliminate sorting overhead with a multi-log}
In this work, we exploit the second key observation to get around the sorting constraint. Namely, at any time only a subset of vertices is being processed from the entire graph. Even using a highly parallel processor the number of vertices that can be handled concurrently will be limited. Hence, only the incoming messages bound for the currently scheduled vertices must be extracted from the log. Thus the process of sorting can be narrowed to a few vertices within the log at a time and can be concurrently performed while a previous batch of vertices is being processed. 

To enable the above capability, PartitionedVC creates a new multi-log structure. Multiple logs are maintained to hold the messages bound for different vertex intervals. We partition the graph into several vertex intervals and associate one log for each interval. As such, we create a coarse-grain log for an interval of vertices that stores all the updates bound to those vertices. As vertices place outbound messages to destination vertices, the destination vertex interval is used as an index to place the message in that log. 

We choose the size of a vertex interval such that typically the entire update log corresponding to that interval can be loaded into the host memory, and used for processing by the vertices in that interval. As the entire update log to a vertex interval can be fitted into the host memory, PartitionedVC avoids the costly step of performing external sort to group the updates bound to a vertex.
\vspace{-2mm}
\subsection{Reduce read overhead with an edge log}
When a vertex interval is processed using a multi-log, one has to still read the outgoing edge information from the CSR so that the processed updates can be placed on the out-edges for processing in the next superstep. The third key observation is that the process of reading the out-edges for each vertex in a vertex interval leads to read amplification. Since the minimum read granularity of an SSD is a page (typically 8-16KBs), one has to read the entire page of an edge list vector that stores the outgoing edges of a particular vertex. Given that many real world graphs exhibit power-law distribution, the vast majority of SSD pages contain the out-edges of multiple vertices. Hence, to read the outgoing edges of a single active vertex an entire page must be fetched.

We measured the fraction of edge list page data that is actually necessary to process a vertex. The data is shown in Figure~\ref{PartitionedVC:fig:edge_list_motivation}. As can be seen from this figure, nearly 32\% of the accessed pages have greater than 0\% and less than 10\% of data activity which is necessary for processing. The inefficient use of a page data leads to significant read amplification, in terms of wasted fetch bandwidth as well as power consumption to move the data. To tackle this challenge, we design an innovative scheme that places all the outgoing edges of any vertex in a separate edge log. We initiate the out-edge logging process while processing an active vertex in the current superstep if that vertex is \textit{likely} to be active again in the next superstep. Thus, the edge log contains the outgoing edges of all likely active vertices to drastically reduce the read amplification overhead. 

Note that whether a vertex is active or not in the next superstep is clearly known if there is a message bound for that vertex in the current superstep. Hence, as vertex intervals are processed the active vertex list for next superstep becomes increasingly obvious. However, in the early part of a superstep we need to predict whether a vertex is likely to be active. We will describe the prediction process in the next section.    

\vspace{-2mm}
\section{Multi-log Architecture}
\label{PartitionedVC:sec:architecture}

\begin{algorithm}[t]
\begin{algorithmic}[1]
\For {all vertex intervals}
\State{$U_{log}$ = LoadLog($V_{in}$)) //Load vertex interval updates}
\State{$S_{log}$ = SortNGrpLog($U_{log}$) //Group updates on vertex\#}
\State{A = ExtractActiveVert($S_{log}$, A)}
\For {each vertex $V_{act}$ in A}
\State{$V_{inf}$=LoadVertexInfo($V_{act}$) // Get $V_{act}$ edges}
\State{$V_{actUpdates}$ = ExtractUpdates($S_{log}$, $V_{act}$)//extract updates bound for $V_{act}$ from $S_{log}$}
\State{ProcessVertex($V_{act}$,$V_{inf}$,$V_{actUpdates}$)}
\EndFor
\EndFor

\Function{SendUpdate}{(v, m)}
\State{$vint_{i}$ = vId2IntervalMap(v)}
\State{Get $vint_{i}$'s top page in the log buffer}
\State{Append update m to the top page}
\State{Flush a page to the SSD on log overflow}
\EndFunction

\end{algorithmic}
\caption{Overview of a superstep in PartitionedVC}
\label{fig:PartitionedVC_framework}
\end{algorithm}

\begin{algorithm}[t]
\begin{algorithmic}[1]
\Function{\textbf{ProcessVertex}}{VertexId $V_{act}$, VertexData $V_{inf}$, VertexUpdates $V_{actUpdates}$}
\For{each update m in $V_{actUpdates}$}
\State{$V_{inf}$.edge(m.source\_id).set\_label(m.data)}
\EndFor
\State{new\_label $=$ frequent\_label($V_{inf}$.edges\_label)}
\State{old\_label $=$ $V_{inf}$.get\_value()}
\If{old\_label $!=$ new\_label}
\State{$V_{inf}$.set\_value(new\_label)}
\For{each edge in $V_{inf}$.edges()}
\State{update m.source\_id = $V_{act}$}
\State{$v_{dest}$ = edge.id(),m.data = new\_label}
\State{SendUpdate($v_{dest}$,m)}
\EndFor
\EndIf
\State{deactivate($V_{act}$)}
\EndFunction
\end{algorithmic}
\caption{Code snippet of community detection program}
\label{fig:FLP_program}
\end{algorithm}

\begin{figure*}[ht]
	\centering
	\includegraphics[width=1\linewidth]{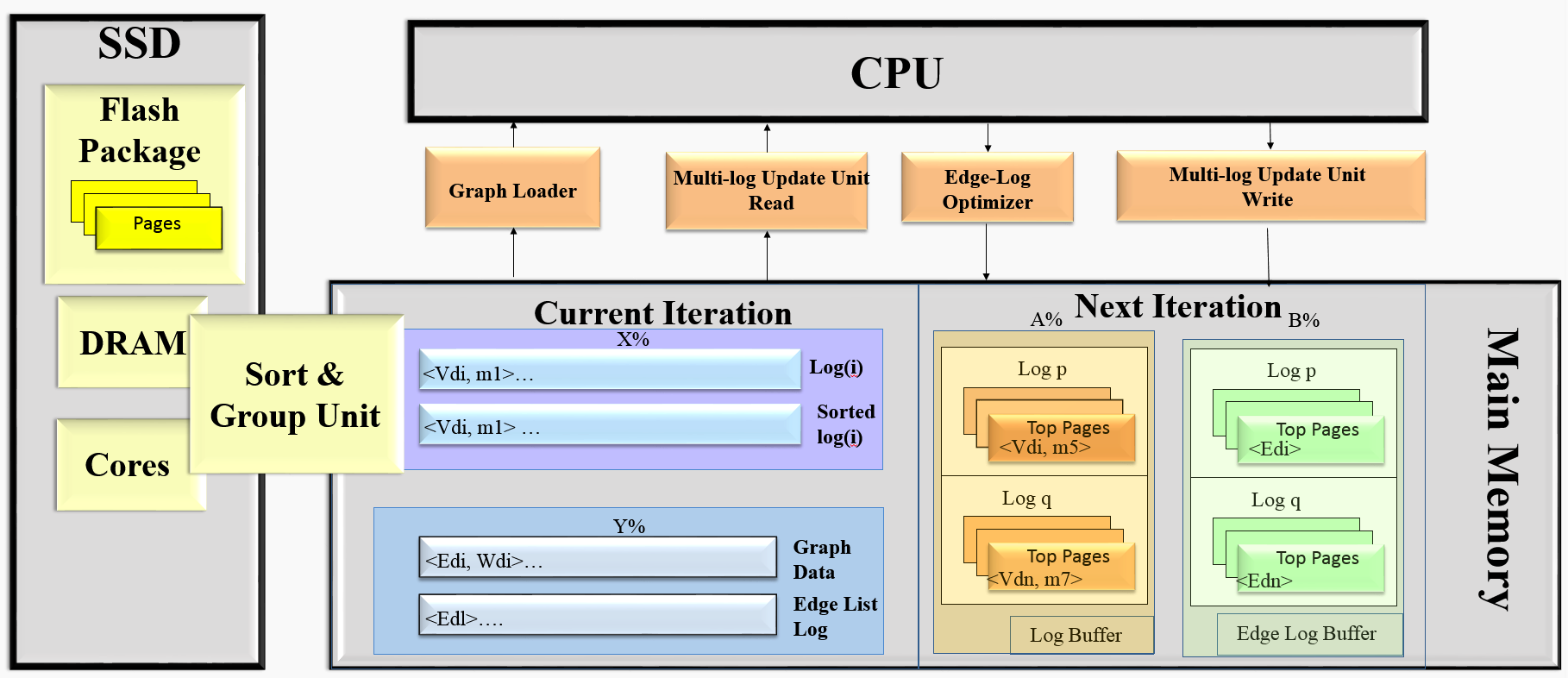}
	\vskip -1mm
	\caption{Layout of PartitionedVC Framework}
	\vskip -5mm
	\label{PartitionedVC:fig:software_components}
\end{figure*}

In this section, we describe the design and implementation details of the PartitionedVC architecture. 
There are two components in PartitionedVC: a set of APIs that are used within each superstep to read, write, and sort logs in service of a user-defined graph application. Algorithm \ref{fig:PartitionedVC_framework} shows the pseudo-code of a single superstep actions in PartitionedVC and the various APIs that activate PartitionedVC functionality. The second component is the memory buffers that store various logs that are accessed by the APIs. Figure \ref{PartitionedVC:fig:software_components} shows these modules as highlighted boxes and their interactions with existing OS and SSD functions. An example code snippet for community detection graph application is shown in \ref{fig:FLP_program}, which uses $ProcessVertex()$ function to provide the vertex-centric computation related to the application.

\vspace{-2mm}
\subsection{Multi-Log Update Unit} 
The multi-log update unit is responsible for managing the logs associated with each vertex interval. Its functionality is activated whenever the $ProcessVertex()$ uses $SendUpdate()$ to send message $m$  to a destination $v_{dest}$. SendUpdate first generates a vertex interval $vint_{i}$ for the vertex $v_{dest}$ using vId2IntervalMap function. It then appends the message $m$ to the log $log_{i}$ associated with that vertex interval. Each message appended to the log is of the format $<v_{dest}, m>$. 

\subsubsection{Generating vertex interval}
Typically, in vertex-centric programming updates are sent over the outgoing edges of a vertex, so the number of updates received by a vertex is at most the number of incoming edges for that vertex. Recall that all updates sent to a single vertex interval are sorted and grouped in the next superstep using in-memory sorting process. Hence, the need to fit the updates of each vertex interval in main memory determines the size of the vertex interval. PartitionedVC conservatively assumes that there may be an update on each incoming edge of a vertex to determine the vertex interval size. It statically partitions the vertices into contiguous segments of vertices, such that the sum of the number of incoming updates to the vertices is less than the memory allocated for the sorting and grouping process. This memory size could be limited by the administrators, application programmer, or could be limited by the size of the virtual machines allocated for graph processing. In our implementation, it is limited to a fraction (shown as X\% in Figure \ref{PartitionedVC:fig:software_components}, which is set to 75\% as the default value) of the total available memory in our virtual machine (1GB is the default).  

\subsubsection{Fusing vertex intervals}
Due to the conservative assumption that there may be a message on each incoming edge, as supersteps progress, the number of messages bound to a vertex interval decrease, and their total size may be less than the allocated memory. To improve memory usage efficiency, PartitionedVC's runtime may dynamically fuse contiguous vertex intervals into a single large interval to process them at once, as we describe later. To enable fusing, PartitionedVC maintains a count of the messages received at each vertex interval to estimate the size of each vertex interval log.  

\subsubsection{SSD-centric log optimizations}
To efficiently implement logging, PartitionedVC first caches log writes in main memory buffers, called the multi-log memory buffers. Buffering helps to reduce fine-grained writes, which in turn reduces write amplification to the SSD storage. PartitionedVC maintains memory buffers in chunks of the SSD page size. Since any vertex interval may generate an update to a target vertex that may be present in any other vertex interval, at least one log buffer is allocated for each vertex interval in the entire graph. In our experiments, even with the largest graph size, the number of vertex intervals was in the order of a few ($<5$) thousands. Hence, at least several thousands of pages may be allocated in multi-log memory buffer at one time. This log buffer is shown to occupy A\% in Figure \ref{PartitionedVC:fig:software_components}, which is about 5\% of the total memory.

For each vertex interval log, a top page is maintained in the buffer. When a new update is sent to the multi-log unit, first the top page of that vertex interval where the update is bound for is identified. As updates are just appended to the log, an update that can fit in the available space on the top page is written into it. If there is not enough space on the top page, then a new page is allocated by the multi-log unit, and that new page becomes the top page for that vertex interval log. PartitionedVC maintains a simple mapping table indexed by the vertex interval to identify the top page.  
    
When the available free space in the multi-log buffer is less than a certain threshold, some log pages are evicted from the main memory to SSD. When a log is evicted from memory it is appended to the corresponding vertex interval log file. Due to the highly concurrent processing of vertices, multiple vertex logs may receive updates and multiple log page evictions may occur concurrently. To maximize log writeback bandwidth, PartitionedVC spans multiple logs across all available SSD channels. Further, each log is interspersed across multiple channels to maximize the read bandwidth when that log file is read back later. We exploit the programmability of the SSD driver in our experimental setup to maximize the write and read bandwidths.  As we make page granular evictions, most of the SSD bandwidth can be utilized. Even when simultaneously writing to multiple logs, they can be parallelized and pipelined due to the interspersing of logs across many channels. Furthermore, we only need to buffer 1000 of SSD pages (each corresponding to a log), our host-side buffer size is also small (in 10-100s of MBs). The reduced main memory usage with multiple logs enables us to allocate a significant fraction of the total memory for fetching, sorting and grouping the updates for processing the superstep, rather than storing the updates for the next superstep. 

\vspace{-2mm}   
\subsection{Sort and Group Unit} 
The sort and group unit is responsible for retrieving the updates from logs and sending the update for processing. Its functionality is automatically activated at the start of each superstep within the PartitionedVC runtime. The first API call is the $LoadLog()$ function, which retrieves the log associated given a vertex interval. The loading process maximizes the SSD read bandwidth since each log is dispersed across all the available flash channels in SSD. After loading a log, if the size of the log is smaller than the memory allocated for sort the next vertex interval log is automatically fetched. Recall that each vertex interval log maintains a counter which provides a first order approximation of the log size in that interval. Hence, the loading process uses this approximate size to determine whether there is enough available memory to fetch the next interval log. The process continues until the memory allocated for sorting is full. The fused logs in memory are then sorted based on the $v_{dest}$ field in each log update.  

\subsubsection{SSD pre-sorting capability} 
During the log updates process, each full page of a vertex interval will be written back to SSD. During the write-back process, the SSD controller can optionally pre-sort each page as it writes the page to the SSD. Since the SSD controller is relatively wimpy, it may not be possible to pre-sort a page when many pages are being written back concurrently. In this case, the SSD controller can skip the pre-sort and directly write the page back to flash. To identify a page that is sorted, a single meta bit in the page is used to indicate whether it is pre-sorted during the writing process.

Note that page-level pre-sorting does not eliminate the need for sorting a vertex interval log. 
The updates bound to a given vertex $v_{dest}$ are dispersed throughout the log associated with that vertex interval. Hence, even if each page in the interval is pre-sorted, it is necessary to sort across pages. However, pre-sorting can reduce the cost of sorting when reading back the log. The $SortNGrpLog()$ function in the PartitionedVC runtime is responsible for this sorting process. 

\subsubsection{Active vertex extraction} 
Once the sorting process is complete, every vertex $V_{dest}$ that has at least one incoming message becomes an active vertex. Hence, the list of $V_{dest}$ from each log is extracted to update the active vertex set $A$ using $ExtractActiveVert()$ API. Then the next superstep only needs to process the active vertices in $A$ by calling the $ProcessVertex()$ function. To process a vertex, any graph algorithm would need the adjacency list (typically the outgoing edges) of that vertex. 

\subsubsection{Graph Loader Unit}
PartitionedVC uses CSR format to store graphs since CSR is more efficient for loading a collection of active vertices. A graph loader unit is responsible to load the graph data for the vertices present in the active vertex list. Graph data unit maintains the row buffer for loading the row pointer and buffer for each of the vertex data (adjacency edge lists/weights). The graph data unit loops over the row pointer array for the range of vertices in the active vertex list, each time fetching vertices that can fit in the graph data row pointer buffer. 
For the vertices which are active in the row pointer buffer, vertex data required by the application such as out-edges or in-edges, are fetched from the \textit{colIdx} or \textit{val} vectors stored in the SSD, accessing only the pages in SSD that have active vertex data.
\vspace{-2mm}   
\subsection{Edge-log optimizer}
The last component of PartitionedVC is the edge-log optimizer module. As we quantified earlier, only a small fraction of the graph pages must be read to fetch the outgoing edges of an active vertex. 
To reduce this read amplification for such pages, PartitionedVC relies on an edge-log optimizer, which works as follows. At the start of the first superstep, the edge log optimizer monitors each vertex $v_i$  that is being processed. Recall that while processing a vertex $v_i$ all its outgoing edges $out_i$ are fetched. First, the edge log optimizer predicts whether $v_i$ will likely be active in the next superstep. An inactive vertex will be activated in the next superstep if it receives a message on its incoming edge. 
If a vertex has not received any incoming message yet, the edge log optimizer has to predict the likelihood that $v_i$ will be active; essentially predicting whether $v_i$ will receive messages on the incoming edges. 

To predict an active vertex, the edge log optimizer uses the past history of active vertices, which are maintained using the active vertex bit vectors. If the current vertex was active at least once in the past $N$ supersteps, it predicts the vertex to be active. More complex prediction schemes were considered, but this simple history-based prediction with $N$ equal to one proved to be effective. 

Once a vertex is predicted to be active, the second step is to determine whether its outgoing edges are located on a page with very few other active vertices. Since the active vertices are being added dynamically, it is not feasible to accurately determine the page usage efficiency. Hence, the edge log optimizer predicts the page usage efficiency for the next superstep based on the current superstep usage efficiency. Pages which use less than a threshold in the current superstep will be predicted as inefficiently used pages. 

Finally, the edge log optimizer logs all the outgoing edges $out_i$ for a predicted active vertex with inefficient page usage into an \textit{edge-log}, indexed with the $v_i$. The edge-log appends the out-edges of active vertices into a sequential page location. Even if a few predictions are incorrect most of the page data contain out-edges of many active vertices, thereby improving read efficiency.  In the next superstep, instead of accessing out-edges from the graph, they can be fetched from the edge-log. In Figure \ref{PartitionedVC:fig:software_components}, this log buffer is set to occupy B\% of the total memory, which is set to 5\% as the default value. 


Note that unlike message logs, edge logs essentially replicate some of the original graph data. 
However, by selecting the thresholds appropriately, graph data replication can be limited. In our implementation, we chose a threshold of 10\% for determining whether a page is efficiently used. With such a low threshold, the critical observation is that when logging $N$ active vertex outgoing edges into a single edge-log page, one can reduce $N-1$ page reads from the original graph.    
\vspace{-2mm}
\subsection{Supporting the generality of vertex-centric programming}
As described in the background section, one of the salient features of PartitionedVC is its ability to support the full spectrum of vertex-centric programming applications. By default, all the messages in the multi-logs are preserved as is, and by using vertex-interval based multi-logs, we can perform in-memory sorting of each of these logs.  

However, some graph algorithms support associative and commutative property on updates. Hence, the updates may be merged into a single update message. PartitionedVC provides an optimization path for such algorithms. For these algorithms, the application developer has to provide the \textit{combine} operator, which may reduce the updates bound to a destination vertex, along with the vertex processing function. The combine operater is applied to all the updates to a target vertex in a superstep before the target vertex's processing function is called. For example, Algorithm \ref{fig:Pagerank_program} shows how the combine function is specified for the pagerank application. When a combine function is defined, the sort and group unit can optimize the performance automatically by preforming the reduction transparently to the user. 

\vspace{-2mm}
\subsection{Graph structural updates:}
\label{subsec:graph_structural_updates}
In vertex-centric programming, graph structure can be updated during the supersteps. Graph structure updates in a superstep can be applied at the end of the superstep. In the CSR format, merging the graph structural updates into the column index or value vectors is a costly operation, as one needs to re-shuffle the entire column vectors. To minimize the costly merging operation, we partition the CSR format graph based on the vertex intervals, so that each vertex interval's graph data is stored separately in the CSR format.

Instead of merging each update directly into the vertex interval's graph data, we batch several structural updates for a vertex interval and later merge them into the graph data after a certain threshold number of structural updates.  As graph structural updates generated during the vertex processing can be targeted to any vertex, we buffer each vertex interval's structural updates in memory. The Graph Loader unit always accesses these buffered updates to fetch the most current graph data for processing accurately.

\vspace{-2mm}
\subsection{Programming model}
\label{subsec:programming_model}
We keep the programming model to be consistent with any vertex-centric programming framework. For each vertex, the vertex processing function is provided. The main function logic for a vertex is written in this function. In the current PartitionedVC implementation the processing function receives the vertex id including the vertex data (such as vertex value), the list of incoming messages for that vertex, and the vertex adjacency information (in all our applications we need the outgoing edges).  Each vertex processes the incoming messages, sends updates to other vertices, and mutate the graph in some applications. Communication between the vertices is implemented using $SendUpdate()$ function, which automatically calls the multi-log update unit transparent to the application developer. The vertex also indicates in the vertex processing function if it wants to be deactivated. If a vertex is deactivated, it will be re-activated automatically if it receives an update from any other vertex. By using multiple logs associated with each vertex interval, PartitionedVC preserves the messages while still enabling in-memory sorting of messages by the sort and group unit. This approach enables PartitionedVC to support the generality of vertex-centric programming model, while reducing the sorting and very large log management overheads.

As described earlier, PartitionedVC provides an optional optimization path if the application developer wants to perform a reduction operation on the incoming messages. For the synchronous computation model, updates will be delivered to the target vertex by the start of its vertex processing in next superstep. In the asynchronous computation model, the latest updates from the source vertices will be delivered to the target vertices, which can either be from the current superstep or the previous one. Vertices can modify the graph structure, but PartitionedVC requires that these graph modifications be finished by the start of next superstep (which is also a restriction placed on most vertex-centric programming models). 






\begin{algorithm}[t]
\begin{algorithmic}[1]
\Function{Combine}{VertexValue val, update m}  
\State{val.change $+=$ m.data.change}
\If{is\_set(m.data.activate)}
\State{activate(m.target\_id)}
\EndIf
\EndFunction
\Function{ProcessVertex}{VertexId $V_{act}$, VertexData $V_{inf}$, VertexUpdates $V_{actUpdates}$}
\For{each edge in $V_{inf}$.edges()}
\State{update $v_{dest}$ = edge.id()}
\State{m.data = $V_{inf}$.val.change$/$$V_{inf}$.num\_edges()}
\If{$V_{inf}$.val.change $>$ Threshold}
\State{m.data.activate = 1}
\EndIf
\State{SendUpdate($v_{dest}$,m)}
\EndFor
\State{value val.page\_rank = ($(1-\alpha)\times$change)}
\State{val.change = 0}
\State{$V_{inf}$.set\_value(val)}
\State{deactivate($V_{act}$)}
\EndFunction
\end{algorithmic}
\caption{Code snippet of Pagerank - an associative and commutative program}
\label{fig:Pagerank_program}

\end{algorithm}

\vspace{-2mm}
\section{System design and Implementation}
\label{PartitionedVC:sec:implementation}
We implemented the PartitionedVC system as a graph analytics runtime on an Intel i7-4790 CPU running at 4 GHz and 16 GB DDR3 DRAM.   We use $2$ TB $860$ EVO SSDs \cite{SSD860EVO}. We use Ubuntu $14.04$ operating system which runs on Linux Kernel version 3.19. 

To simultaneously load pages from several non-contiguous locations in SSD using minimum host side resources, we use asynchronous kernel IO. To match SSD page size and load data efficiently, we perform all the IO accesses in granularities of 16KB, typical SSD page size \cite{ssd_page_size}. Note that the load granularity can be increased easily to keep up with future SSD configurations. The SSD page size may keep growing to accommodate higher capacities and IO speeds, SSD vendors are packing more bits in a cell to increase the density, which leads to higher SSD page sizes \cite{enterprise_storage}.

We used OpenMP to parallelize the code for running on multi-cores. We use an 8-byte data type for the rowPtr vector and 4 bytes for the vertex id. Locks were sparingly used as necessary to synchronize between the threads. With our implementation, our system can achieve 80\% of the peak bandwidth between the storage and host system.

\textbf{Baseline:} We compare our results with the popular out-of-core GraphChi framework. While several recent works have optimized GraphChi \cite{vora2016load,zhang2018wonderland,ai2018clip}, most of these works have focused on optimizing algorithms that satisfy the commutative and associative property on their updates. Hence, these approaches tradeoff the generality of GraphChi for improving performance. Since PartitionedVC strives to preserve the generality, we believe GraphChi is the most appropriate comparison framework. On the log-based execution front, we compare our approach with GraFBoost \cite{GraFBoost}. GraFBoost uses a single log to maintain all updates, and exploits the commutative and associative property of graph algorithms to merge the updates to shorten the log. The merging process enables them to perform a relatively efficient out-of-memory sorting. Due to the limitation of GraFBoost, we can only compare PartitionedVC when the algorithms satisfy the constraints of GraFBoost.   

While comparing with GraphChi, we use the same host-side memory cache size as the size of the multi-log buffer used in PartitionedVC. In both our implementation and GraphChi's implementation, we limit the memory usage to $1$ GB, primarily because the real-world graph datasets that are available are at most 100 GBs, and this approach has been used in many prior works to emulate a realistic memory-graph size ratio \cite{matam2019graphssd, zhang2018wonderland, ai2018clip}.  
In our implementation, we define memory usage by limiting the total size of the multi-log buffer. GraphChi provides an option to specify the amount of memory budget that it can use. We maximized GraphChi performance by enabling multiple auxiliary threads that GraphChi may launch. As such, GraphChi also achieves peak storage access bandwidth. 



\textbf{Graph dataset:}
To evaluate the performance of PartitionedVC, we selected two real-world datasets, one from the popular SNAP dataset~\cite{snapnets}, and the other one is a popular web graph from Yahoo Webscope dataset \cite{yahooWebscopre_graph}. These graphs are all undirected graphs, and for an edge, each of its end vertices appears in the neighboring list of the other end vertex. Table \ref{PartitionedVC:table:graph_dataset} shows the number of vertices and edges for these graphs.

\begin{table}[ht]
\centering
\scalebox{0.90} {
 \begin{tabular}{|c | c | c |} \hline
Dataset name    & Number of vertices & Number of edges \\ \hline
com-friendster (CF) &    124,836,180  &    3,612,134,270\\ \hline
YahooWebScope (YWS) &    1,413,511,394  &    12,869,122,070\\ \hline
\end{tabular}
}
\caption{Graph dataset}
\vskip -0.3cm
\label{PartitionedVC:table:graph_dataset}
\end{table}

\vspace{-3mm}
\section{Applications}
\label{PartitionedVC:sec:applications}
To illustrate the benefits of our framework, we evaluate two classes of graph applications. The first set of algorithms allows updates to be merged with no impact on correctness, which are proper workloads for GraFBoost. The second set requires all the updates to be individually handled, which can only be evaluated on GraphChi and PartitionedVC.

\textbf{Merging updates acceptable:} BFS, Pagerank (PR)\cite{pagerank_implementation}. GraFBoost works well with these algorithms. 

\textbf{Merging updates not possible:} Community detection (CD)\cite{FLP_implementation}, Graph coloring (GC)\cite{gonzalez2012powergraph}, Maximal independent set (MIS)\cite{salihoglu2014optimizing}, Random walk (RW) \cite{kyrola2013drunkardmob}, and K-core\cite{Quick:2012:UPL:2456719.2457085}.  All the applications are implemented using the details presented in the references. 

\textbf{Additional application details:} 
A vertex in page rank gets activated if it receives a delta update greater than a certain threshold value (0.4). For random walk, we sampled every $1000^{th}$ node as a source node and performed a random walk for 10 iterations with a maximum step size of 10.

Due to extremely high computational load, for all the applications we ran 15 supersteps or less than that if the problem converges before that. Many prior graph analytics systems also evaluate their approach by limiting the superstep count~\cite{GraFBoost}. 
All the applications mentioned above can be implemented using GraphChi and MulilogVC. However, GraFBoost can only work for associative and commutative applications (Pagerank, BFS).
\vspace{-2mm}
\section{Experimental evaluation}
\label{PartitionedVC:sec:Evaluation}
\begin{figure*}[h]
    \centering
    \subfloat[BFS relative to GraphChi]{
        \includegraphics[width=0.34\linewidth,height=4cm]{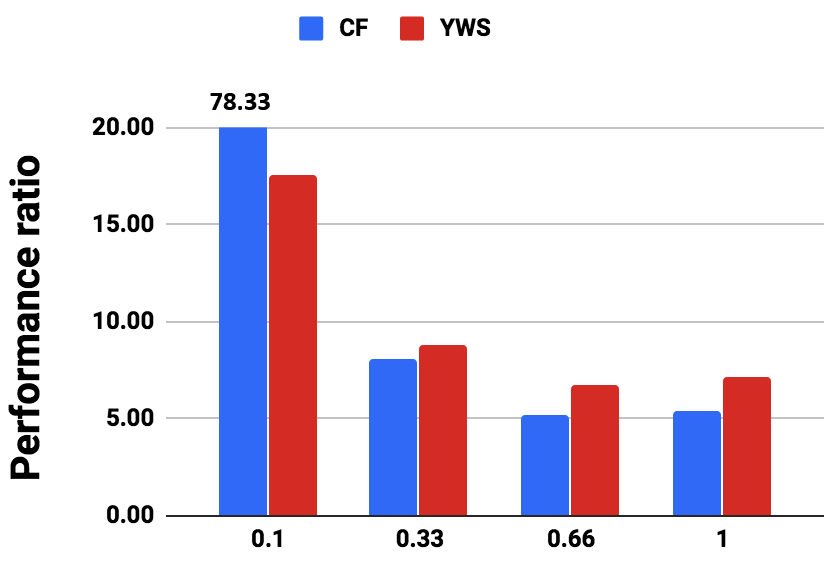}
        \label{fig:graphchi_bfs_comparison}
    }
    \subfloat[Page access counts ratio relative to PartitionedVC]{
        \includegraphics[width=0.34\linewidth,height=4cm]{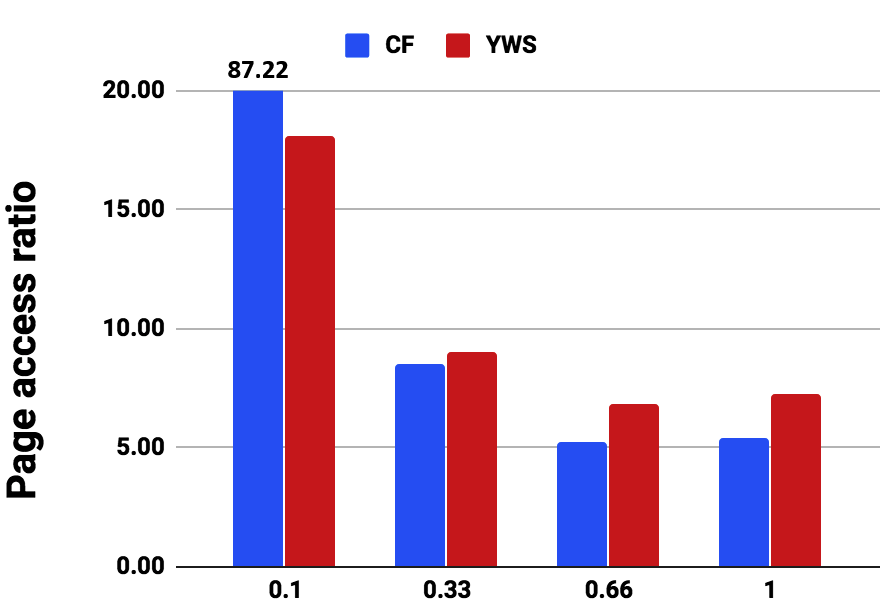}
        \label{fig:bfs_page_access}
    }
    \subfloat[Execution Time Breakdown]{
        \includegraphics[width=0.3\linewidth,height=4cm]{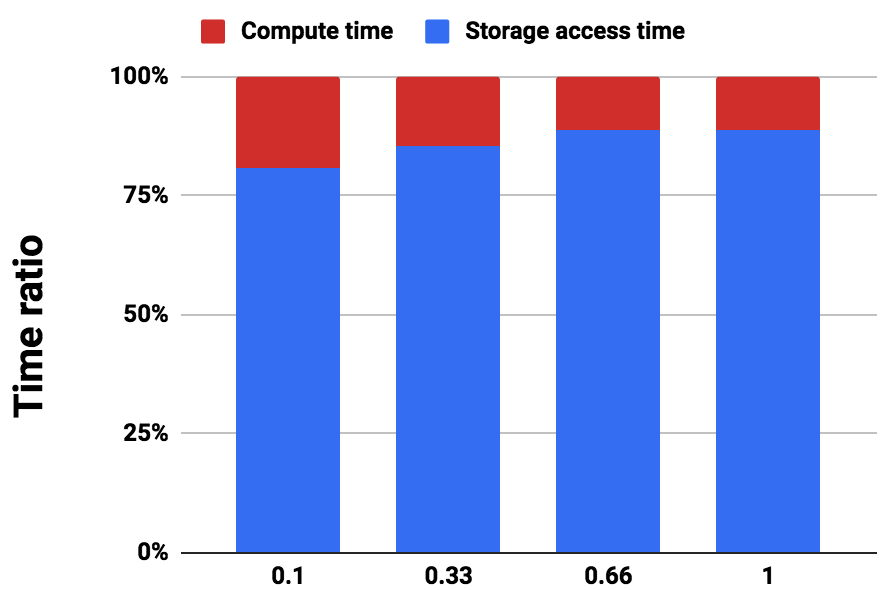}
        \label{fig:bfs_time_access}
    }
    \vskip -1mm
    \caption{BFS Application Performance}
    \vskip -4mm
    \label{PartitionedVC:fig:results_baseline}
\end{figure*}

Figure \ref{fig:graphchi_bfs_comparison} shows the performance comparison of BFS application on our PartitionedVC and GraphChi frameworks. The X-axis shows the selection of a target node that is reachable from a given source by traversing a fraction of the total graph size. Hence, an X-axis of 0.1 means that the selected source-target pair in BFS requires traversing 10\% of the total graph before the target is reached. We ran BFS with different traversal demands. The Y-axis indicates speedup which is the  application execution time on GraphChi divided by application execution time on PartitionedVC framework. 

On average BFS performs $17.8X$  better on PartitionedVC when compared to GraphChi. Performance benefits come from the fact that PartitionedVC accesses only the required graph pages from storage.  BFS has a unique access pattern. Initially, as the search starts from the source node it keeps widening. Consequently, the size of the graph accessed and correspondingly the update log size grows after each superstep. As such, the performance of PartitionedVC is much higher in the initial supersteps and then reduces in later intervals. 

Figure~\ref{fig:bfs_page_access} show the ratio of page accesses in GraphChi divided by the page accesses in PartitionedVC. GraphChi loads nearly 90X more data when using 0.1 (10\%) traversals. However, as the traversal need increases, GraphChi loads only 6X more pages. As such, the performance improvements seen in BFS are much higher with PartitionedVC when only a small fraction of the graph needs to be traversed. Figure~\ref{fig:bfs_time_access} shows the distribution of the total execution time split between storage access time (which is the load time to fetch all the active vertices) and the compute time to process these vertices. The data shows that when there is a smaller fraction of the graph that must be traversed, the storage access time is about 75\%, however as the traversal demands increase the storage access time reaches nearly 90\% even with PartitionedVC. Note that with GraphChi the storage access time stays nearly constant at over 95\% of the total execution time.

\begin{figure*}[h]
    \centering
    \subfloat[Pagerank]{
        \includegraphics[width=0.18\linewidth,height=3cm]{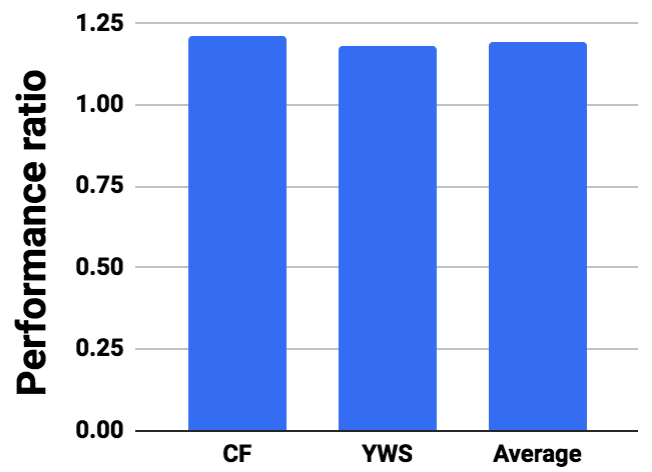}
        \label{fig:graphchi_pagerank_comparison}
    }
    \subfloat[FLP]{
        \includegraphics[width=0.18\linewidth,height=3cm]{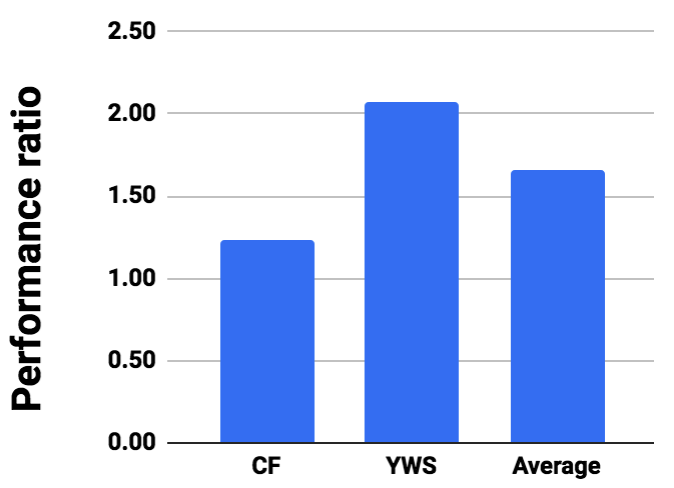}
        \label{fig:graphchi_flp_comparison}
    }
    \subfloat[GC]{
        \includegraphics[width=0.18\linewidth, height=3cm]{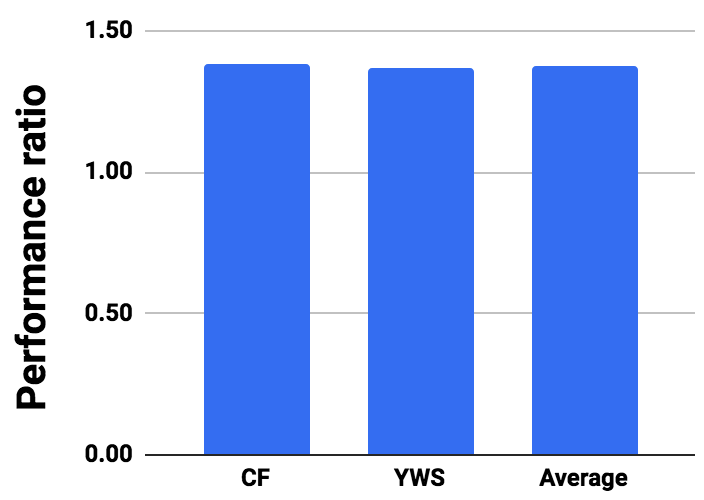}
        \label{fig:graphchi_GGC_comparison}
    }
    \subfloat[MIS]{
        \includegraphics[width=0.18\linewidth, height=3cm]{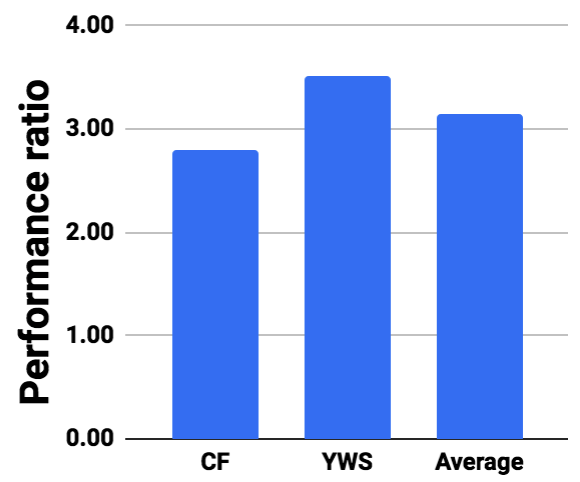}
        \label{fig:graphchi_MIS_comparison}
    }
    \subfloat[RW]{
        \includegraphics[width=0.18\linewidth, height=3cm]{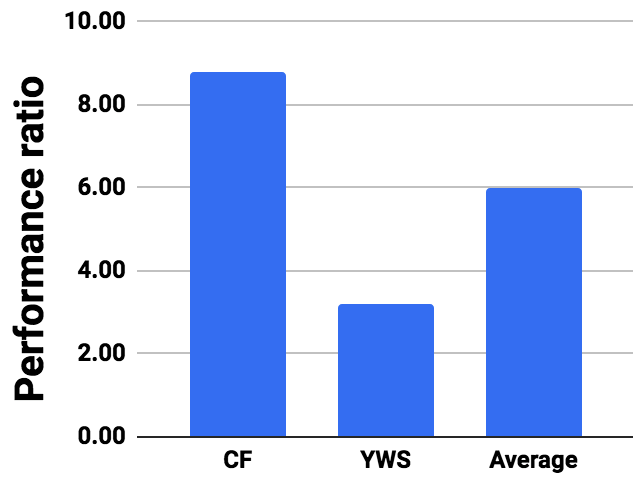}
        \label{fig:graphchi_RW_comparison}
    }
    \vskip -2mm
    \caption{Application performance relative to GraphChi}
    \vskip -4mm
    \label{fig:results_baseline_extended}
\end{figure*}

\begin{figure*}[h]
    \centering
    \subfloat[Pagerank]{
        \includegraphics[width=0.2\linewidth,height=4cm]{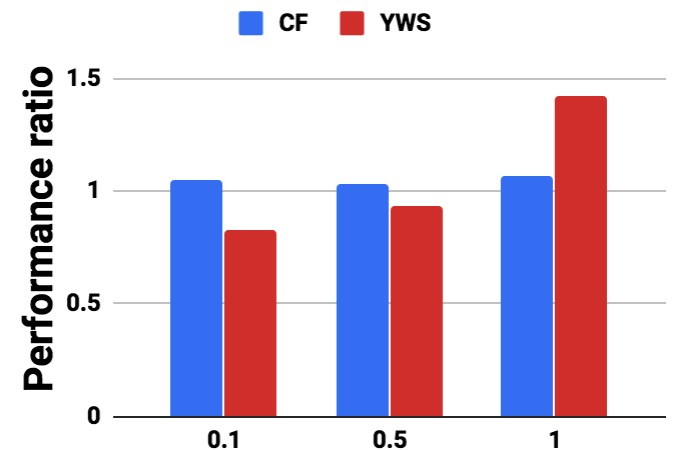}
        \label{fig:pagerank_iteration_comparison}
    }
    \subfloat[FLP]{
        \includegraphics[width=0.2\linewidth,height=4cm]{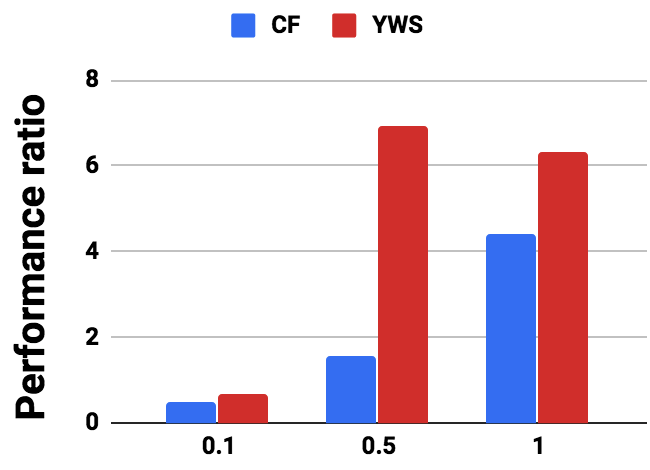}
        \label{fig:flp_iteration_comparison}
    }
    \subfloat[GC]{
        \includegraphics[width=0.2\linewidth,height=4cm]{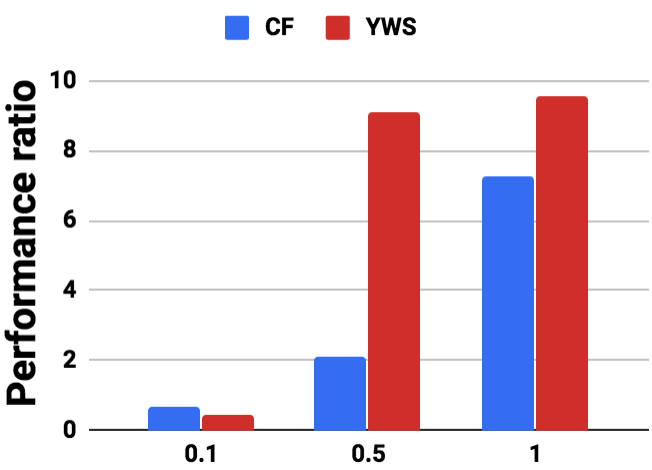}
        \label{fig:GGC_iteration_comparison}
    }
    \subfloat[MIS]{
        \includegraphics[width=0.2\linewidth,height=4cm]{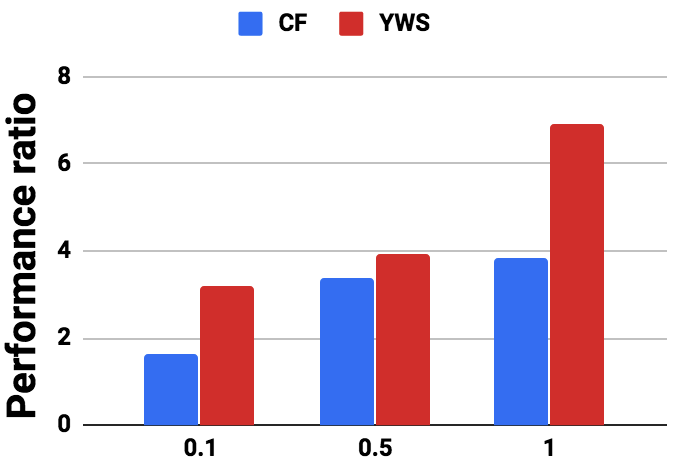}
        \label{fig:MIS_iteration_comparison}
        }
        \subfloat[RW]{
        \includegraphics[width=0.2\linewidth,height=4cm]{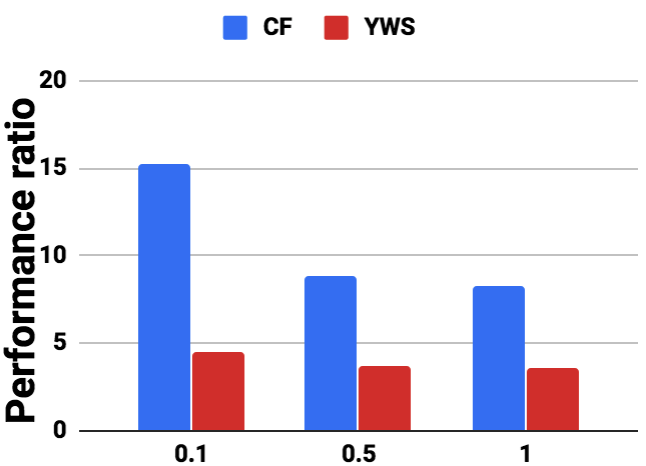}
        \label{fig:RW_iteration_comparison}
        }

    \vskip -2mm
    \caption{Application performance comparisons over supersteps}
    \vskip -4mm
    \label{fig:results_iterations}
\end{figure*}

Figure \ref{fig:graphchi_pagerank_comparison} shows the performance comparison of pagerank. On average, pagerank performs $1.2X$  better with PartitionedVC. Unlike BFS, pagerank has an opposite traversal pattern. In the early supersteps, many of the vertices are active and many updates are generated. But during later supersteps, the number of active vertices reduces and PartitionedVC performs better when compared to GraphChi. Figure \ref{fig:pagerank_iteration_comparison} shows the performance of PartitionedVC compared to GraphChi over several supersteps. Here X-axis shows the superstep number as a fraction of the total executed supersteps. During the first half of the supersteps PartitionedVC has similar, or in the case of YWS dataset slightly worse performance than GraphChi. The reason is that the size of the log generated is large. But as the supersteps progress and the update size decreases the performance of PartitionedVC gets better. 



\begin{figure}[h]
\vskip -2mm
\subfloat[Performance of K-Core algorithm]{
        \includegraphics[width=0.5\linewidth,height=3.5cm]{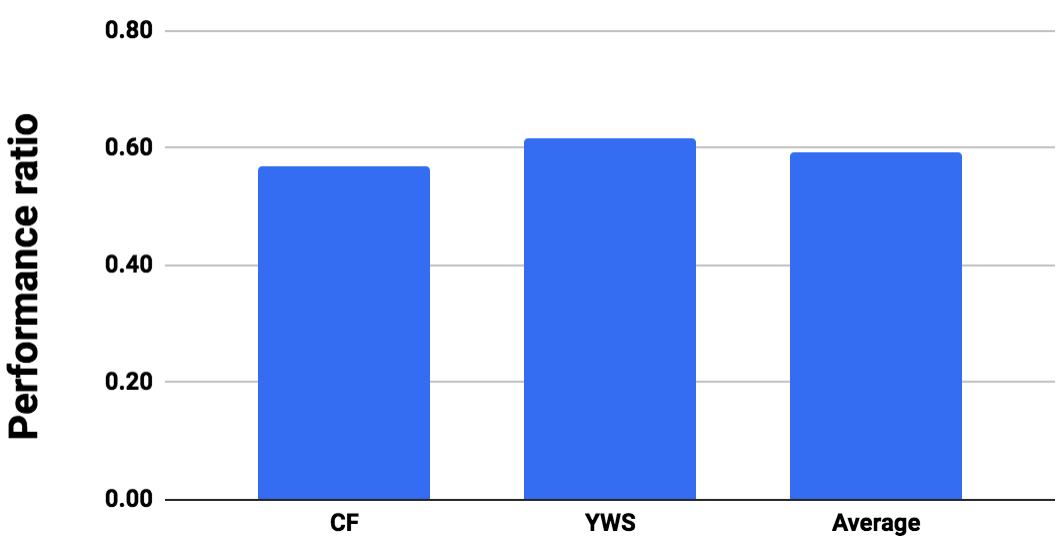}
    \label{fig:KCore_algoritm}
    }
    \subfloat[GraFBoost Performance]{
        \includegraphics[width=0.5\linewidth,height=3.5cm]{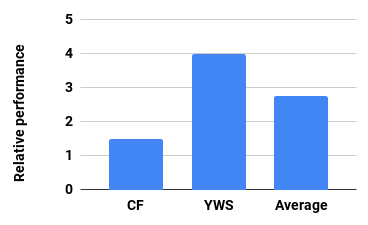}
        \label{fig:graf_boost}
        }
    \vskip -2mm
    \caption{Performance comparison}
    \vskip -8mm
    \label{fig:performance_comparison_kcore_grafboost}
\end{figure}


Figure \ref{fig:graphchi_flp_comparison} shows the performance of community detection (FLP), which is similar in behavior to graph coloring (Figure \ref{fig:graphchi_GGC_comparison}). On average community detection performs $1.7X$ times better on PartitionedVC over GraphChi. Figure \ref{fig:flp_iteration_comparison} shows performance at supersteps. Similar to pagerank application, initially a lot of the vertices are active, and in later supersteps fewer vertices are active. In community detection (and graph coloring) application, target vertices receive updates from source vertices over in-edges. Also active vertices access in-edge weights and store the updates received via source vertices so that a vertex can only send its label if it has been changed. As a result, in PartitionedVC framework, community detection application has to access both updates and edge-weights from storage. Whereas in GraphChi, as updates are passed to the target vertices via edge weights, only edge-weights need to be accessed from storage. In later supersteps, PartitionedVC sees few active vertices, and even with the need to access updates and edge-weights separately, it outperforms GraphChi.


Figure \ref{fig:graphchi_MIS_comparison} shows the performance of maximal independent set algorithm. On average maximal independent set algorithm performs $3.2\times$  better on PartitionedVC when compared to GraphChi. In this algorithm also, as vertices are selected with a probability, there are fewer active vertices in a superstep. Figure \ref{fig:MIS_iteration_comparison} shows the performance over supersteps.

Figure \ref{fig:graphchi_RW_comparison} shows the performance comparison of random walk application on PartitionedVC framework with GraphChi framework. On average random walk is $6\times$  faster than GraphChi.
This algorithm follows the same pattern as BFS. Initially, its active vertices subset is smaller, and gradually it expands and eventually tapers off. Hence, PartitionedVC shows significant improvements in the earlier supersteps and, gradually the performance delta also tapers off.

\textbf{GraFBoost comparison:} Since GraFBoost only works on applications where updates can be merged into a single value, we use pagerank, which satisfies GraFBoost limitation. Figure \ref{fig:graf_boost} presents the performance. The Y-axis illustrates PartitionedVC performance improvement relative to GraFBoost. To have a fair comparison, for both of the systems we use the same configuration of $1 GB$ memory. For GraFBoost, memory usage is limited by CGGROUP instructions. Also, since GraFBoost currently does not support loading only active graph data, the comparison is done based on only first iteration. On average, PartitionedVC is $2.8\times$ faster than GraFBoost. As can be observed from this figure, Yahoo Web dataset which is a larger dataset shows a significant performance improvement ($4$ times faster than GraFBoost). The reason is that as the dataset grows, the log file size also grows. Therefore, the cost of sorting large logs that do not fit in memory dominates in GraFBoost. 

\textbf{Adapting GraFBoost for applications with non-mergeable updates:} We compare the performance of graph coloring application by adapting GraFBoost's single log structure for passing the update messages. As we cannot merge the updates generated to a target vertex into a single value, we need to keep all the updates and sort them. For graph coloring application, when compared to this adapted GraFBoost, our PartitionedVC performs $2.72$x and $2.67$x for CF and YWS, respectively.

PartitionedVC uses CSR format, which enables accessing only active vertices, but graph updates are costly. Using multiple intervals as separate CSR structures certainly help reduce the update cost of CSR format. Figure \ref{fig:KCore_algoritm} shows the performance of K-core, which is the most demanding application for PartitionedVC. K-core uses delete operations that modify the graph. Furthermore, K-core has only one iteration in the algorithm, and all the vertices are active in that iteration. GraphChi can directly update the delete bit in its outgoing edge's shard, whereas in PartitionedVC we log the structural update and later update it in the graph. Even in this pathological case, PartitionedVC achieves about 60\% of the GraphChi performance. 
However, for other structural update operations like add edge or vertex, GraphChi and PartitionedVC tackle the structural updates in a similar fashion; namely, they both buffer the updates and merge after a threshold. 


\begin{figure}
\centering
        \includegraphics[width=\linewidth,height=4cm]{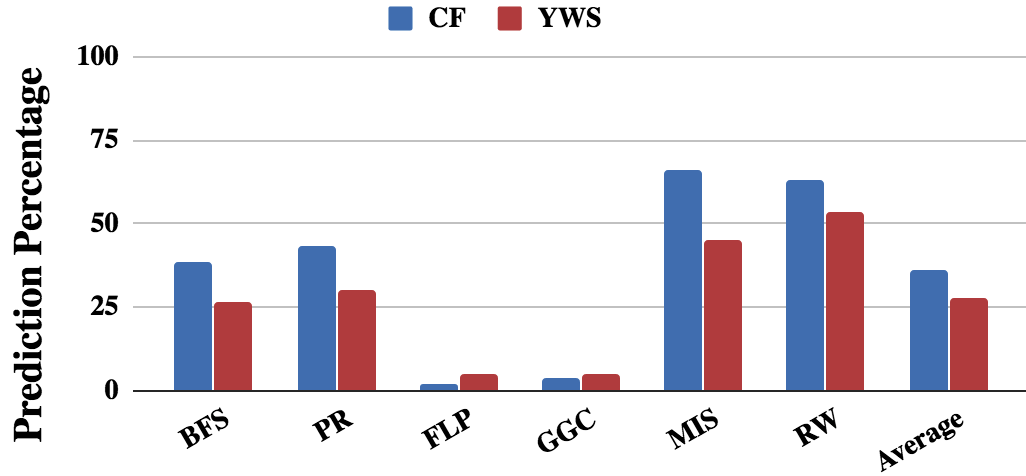}
    \vskip -3mm
    \caption{Percentage of inefficient pages predicted}
    \vskip -8mm
    \label{fig:edge_log_prediction_percentage}
\end{figure}
\textbf{Edge-log optimizer prediction accuracy:} Figure \ref{fig:edge_log_prediction_percentage} shows the percentage of inefficiently used graph pages (i.e. pages with $>$ $0$ and $<$ $10$\% of their content utilized) that we predicted correctly. On average, our scheme predicts $34$\% of the inefficiently used pages. In FLP and GGC applications, as they converge faster, the number of inefficient pages are fewer (shown in Figure \ref{PartitionedVC:fig:edge_list_motivation}) and concomitantly our history-based prediction model predicts with less accuracy. However, for other applications with higher inefficient pages over several iterations, our prediction accuracy is higher.

\begin{figure}
\centering
    \includegraphics[width=0.7\linewidth,height=3.5cm]{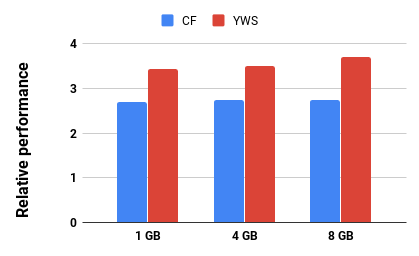}
    \vskip -4mm
    \caption{Memory scalability}
    \vskip -4mm
    \label{fig:memory_scalability}
\end{figure}

\textbf{Memory scalability:} To study the scalability, we conducted experiments with increasing main memory to $4$ GB and $8$ GB. Figure \ref{fig:memory_scalability} shows the performance of PartitionedVC over GraphChi for MIS application. As memory size increases, the relative improvement of  PartitionedVC over GraphChi stays about the same with about 5-10\% increase in the performance of  PartitionedVC when using larger memory. 

\vspace{-2mm}
\section{Related work}
\label{PartitionedVC:sec:RelatedWork}

Graph analytics systems are widely deployed in important domains such as drug processing, social networking, etc., concomitantly there has been a considerable amount of research in graph analytics systems \cite{low2014graphlab,gonzalez2012powergraph, nai2015graphbig}.

For vertex-centric programs with associative and commutative functions, where one can use a combine function to accumulate the vertex updates into a vertex value, GraFBoost implements an external memory graph analytics system \cite{GraFBoost}. It logs updates to the storage and pass vertex updates in a superstep. They use a sort-reduce technique for efficiently combining the updates and applying them to vertex values. 
However, they may access storage multiple times for the updates, as they sort-reduce on a single giant log. In our system, we keep multiple logs for the updates in storage, and in a superstep we access both the graph pages and the update pages once, corresponding to the active vertex list. Also, our PartitionedVC system supports, complete vertex-centric programming model rather than just associative and commutative combine functions, which has better expressiveness and makes computation patterns intuitive for many problems \cite{apache_flink}.

A recent work \cite{elyasi2019large} extends GraFBoost work. 
In this work, they avoid sorting by partitioning the destination vertices such that they can fit in the main memory and replicate the source vertices across the partitions so that when processing a partition, all the source vertices graph data can be streamed in and updates to destination vertex can be performed in the main memory itself. However, one may need to replicate the source vertices and access edge data multiple times. To support complete vertex-centric programming with this scheme, computing may be prohibitively expensive, as the number of partitions may be high, the replication cost will also be high. 



X-stream \cite{roy2013x} and GridGraph \cite{zhu2015gridgraph} are edge-centric based external memory systems which aim to sequentially access the graph data stored in secondary storage. Edge-centric systems provide good performance for programs which require streaming in all the edge data and performing vertex value updates based on them. However, they are inefficient for programs which require sparse accesses to graph data such as BFS, or programs which require access to adjacency lists for specific vertices such as random-walk.

GraphChi \cite{graphchi} is the only external memory based vertex-centric programming system that supports more than associative and commutative combine programs. 
In this work, we compare with GraphChi as a baseline and show considerable performance improvements.

There are several works which extend GraphChi by trying to use all the loaded shard or minimizing the data to load in shard \cite{ai2017squeezing, vora2016load}. However, in this work, we avoid loading data in bulky shards at the first place and access only graph pages for the active vertices in the superstep.

Semi-external memory systems such as Flash graph \cite{flashgraph}, GraphMP \cite{sun2017graphmp} stores the vertex data in the main memory and achieves high performance. When processing with a low-cost system and available main memory is less than the vertex value data, these systems suffer from performance degradation due to the fine-grained accesses to the vertex-value vector.

Due to the popularity of graph processing, they have been developed in a wide variety of system settings. In distributed computing setting, there are popular vertex-centric programming based graph analytic systems including Pegasus \cite{kang2009pegasus}, Pregel \cite{malewicz2010pregel}, Graphlab \cite{low2014graphlab}, Blogel \cite{yan2014blogel}, PowerGraph \cite{tiwari_hotpower12}, Gemini \cite{zhu2016gemini}, Kickstarter \cite{vora2017kickstarter}, Coral~\cite{vora2017coral}. In the single-node in-memory setting there are many frameworks such as Turbograph \cite{han2013turbograph}, Venus \cite{cheng2015venus}, Graphene \cite{liu2017graphene}, Mosaic \cite{maass2017mosaic}, GraphM \cite{zhao2019graphm}. Ligra \cite{shun2013ligra} provides a framework optimized for multi-core processing. Lumos is an out of core graph processing engine that provided a synchronous processing guarantee by using cross iteration value propagation \cite{vora2019lumos}. Polymar \cite{zhang2015numa} is a Numa-aware multicore engine that is designed to optimize the remote access. Some engines such as Galois \cite{nguyen2013lightweight} supports both single machine and distributed systems. 

\vspace{-2mm}
    \section{Conclusion}
    \label{PartitionedVC:sec:Conclusion}
    Graph analytics are at the heart of a broad set of applications. In external-memory based graph processing system, accessing storage becomes the bottleneck. However, existing graph processing systems try to optimize random access reads to storage at the cost of loading many inactive vertices in a graph. In this paper, we use CSR format for graphs that are more amenable for selectively loading only active vertices in each superstep of graph processing. However, CSR format leads to random accesses to the graph during update process. We solve this challenge by using a multi-log update system that logs  updates in several log files, where each log file is associated with a single vertex interval. Further, while accessing SSD pages with fewer active vertex data, we reduce the read amplification due to the page granular accesses in SSD by logging the active vertex data in the current iteration and efficiently reading the log in the next iteration. Over the current state of the art out-of-core graph processing framework, our evaluation results show that PartitionedVC framework improves the performance by up to $17.84\times$, $1.19\times$, $1.65\times$, $1.38\times$, $3.15\times$, and $6.00\times$ for the widely used breadth-first search, pagerank, community detection, graph coloring, maximal independent set, and random-walk applications, respectively.




\end{document}